# Laser Inter-Satellite Links in a Starlink Constellation

Aizaz U. Chaudhry and Halim Yanikomeroglu, Carleton University

*Laser inter-satellite links (LISLs) are envisioned between satellites in upcoming satellite constellations, such as Phase I of SpaceX's Starlink. Within a constellation, satellites can establish LISLs with other satellites in the same orbital plane or in different orbital planes. We present a classification of LISLs based on the location of satellites within a constellation and the duration of LISLs. Then, using satellite constellation for Phase I of Starlink, we study the effect of varying a satellite's LISL range on the number of different types of LISLs it can establish with other satellites. In addition to permanent LISLs, we observe a significant number of temporary LISLs between satellites in crossing orbital planes. Such LISLs can play a vital role in achieving low-latency paths within next-generation optical wireless satellite networks.*

## I. Introduction

Optical wireless communications in outdoor environments is referred to as *free space optics* (FSO). An FSO link is used to transmit an optical signal from an optical transmitter to an optical receiver over the atmosphere or the vacuum in space, and a clear line of sight is required between transmitter and receiver [1]. FSO links can be of four types, namely, terrestrial, non-terrestrial (or aerial), space, and deep-space. Satellite-to-satellite FSO links are examples of space FSO, and are referred to in this work as *laser inter-satellite links* (LISLs).

Higher frequency, higher bandwidth, and the characteristics of their laser beam provide LISLs the following significant advantages over radio frequency-based inter-satellite links: higher data rates; smaller antenna sizes resulting in lesser weight and lesser volume; narrower beams eliminating interference and providing higher security; and lower requirement of transmit power as a result of lower beam spread and higher directivity [2]. LISL terminals can be easily integrated into satellite platforms as they require less onboard satellite resources due to their smaller size, weight, volume, and power requirement. The satellite launching and deployment costs are also reduced due to the smaller form factor of LISL terminals [3].

The high capacity and low latency demands of next-generation satellite networks can be fulfilled by employing LISLs between satellites in a constellation. Otherwise, a long-distance inter-continental connection between two cities, such as New York and Hong Kong, will have to ping-pong between ground stations and satellites, and this will negatively impact network latency. Next-generation satellite networks arising from upcoming satellite constellations, like SpaceX's Starlink, are expected to be fully operational by the mid to late 2020s, and LISLs are envisaged between satellites in these upcoming satellite networks. SpaceX has an ambitious aim to deploy Starlink to deliver high-speed broadband Internet service to users around the world. Starlink is expected to be comprised of approximately 12,000 satellites in different *low Earth orbit* (LEO) and *very low Earth orbit* (VLEO) constellations. In Phase I of Starlink, SpaceX is actively deploying an LEO constellation of 1,584 satellites [4]. LISL terminals are being developed and are expected to offer capacities of up to 10 Gbps [5–7]. To set up an effective global communications network in space, LISLs offering capacities in the hundreds of Gbps – progressively increasing to Tbps over time – will be required.

In this work, we present a classification of LISLs based on the location of satellites within a constellation and the duration of LISLs. Next, we use the satellite constellation for Phase I of Starlink to study the effect of varying a satellite's LISL range on the number of different types of LISLs it can establish with other satellites in the constellation. Besides other type of permanent and temporary LISLs, we find a significant number of temporary LISLs between satellites in crossing orbital planes. Such LISLs can be vital in achieving low-latency paths within next-generation optical wireless satellite networks that are envisioned by the mid to late 2020s.

The rest of the paper is organized as follows: Section II explores related work. Classification of LISLs is discussed in Section III. The results for studying the impact of different satellite LISL ranges on the number of different types of LISLs are presented in Section IV. Conclusions and some possible directions for future work are highlighted in Section V.

## II. Motivation

SpaceX planned to equip its Starlink satellites with five LISLs so that they could connect to other satellites within the constellation to establish a satellite network. This is evident from five 1.5 kg silicon carbide communication components that are mentioned by SpaceX in its 2016 FCC filing [8] while discussing space debris issues when its Starlink satellites reach end of life and need to be de-orbited. Silicon carbide is used in mirrors in LISL terminals. However, SpaceX revised the number of LISLs (i.e., the number of silicon carbide communication components) per satellite to four in a later FCC filing in 2018 [4].

A comparison of laser and radio frequency (RF) links has been conducted when employed between two LEO satellites [3]. Two different RF links operating in Ka and mm-wave bands are examined, and the transmit power is considered as 10 W, 20 W, and 50 W for laser (193 THz), mm-wave band (60 GHz) and Ka band (32 GHz) links, respectively. The RF inter-satellite link in either Ka or mm-wave bands needs at least 19 times the antenna diameter, and more than twice the onboard power and mass compared to the laser inter-satellite link for a link data rate of 2.5 Gbps and an inter-satellite distance of 5,000 km.

LISL terminals will be extremely critical in the formation of a global space communications network by inter-connecting hundreds of satellites via LISLs in next-generation satellite networks. Tesat [5] has developed two such LISL terminals while others are under development by Mynaric [6] and General Atomics [7].

The problem of designing the inter-satellite network for low latency and high capacity has been explored [9]. Repetitive patterns in the network topology, referred to as motifs, were employed to avoid link changes over time. Four LISLs per satellite were assumed, in line with SpaceX's recent FCC filings [4], to connect to two neighbors in the same orbital plane and to two neighbors in two different orbital planes.

How to use LISLs to provide a network within a satellite constellation and the problem of routing on this network has been investigated [10]. It was mentioned that a network built using LISLs could provide lower latency communications than any possible terrestrial optical fiber network for communications over distances greater than about 3,000 km. As per SpaceX's earlier FCC filings [8], it was assumed that each satellite would have five LISLs to connect to other satellites within the constellation. The fifth LISL was used to connect to a satellite in a crossing orbital plane. SpaceX has changed the number of LISLs per satellite to four, and this could be due to the difficulty in developing the hardware capability for this fifth LISL.

The primary use case of next-generation optical wireless satellite networks that are formed from LISLs between satellites may turn out to be the provision of low-latency communications over long distances. By providing low-latency communications as a premium service to the financial hubs around the world, the cost of establishing and maintaining such networks can easily be recovered. In trading stocks at the stock exchange, it is estimated that a 1 millisecond advantage can be worth $100 million a year to a single major brokerage firm [11], and an advantage of a few milliseconds may result in billions of dollars of revenues for these financial firms. To reduce latency, these firms are looking for technological solutions, and low-latency next-generation optical wireless satellite networks may offer the perfect solution.

A use case for next-generation optical wireless satellite networks is investigated to analyze their suitability for low-latency communications over long distances [12]. It is shown that an optical wireless satellite network operating at 550 km altitude outperforms a terrestrial optical fiber network in terms of latency for communication distances of more than 3,000 km.

A study analyzing the effect of varying the satellite LISL range on the number of different types of LISLs a satellite can establish with other satellites in a constellation does not exist in the literature. No previous work has analyzed the number of different types of LISLs that can exist between satellites in a constellation.

## III. Classification of LISLs

A LEO (or VLEO) satellite constellation can have several orbital planes and each orbital plane can have numerous LEO (or VLEO) satellites. LISLs can be classified into two main types based on the location of satellites within a constellation: *intra-orbital plane LISL*, which is established between two satellites in the same orbital plane; and *inter-orbital plane LISL*, which is created between satellites in two different orbital planes. Within the same *orbital plane* (OP), satellites at the same altitude move in the same direction, which means that these satellites move with the same velocity.

Inter-orbital plane LISLs can be further divided into three types: *adjacent OP LISL* (AOPL), which can be formed between satellites in adjacent orbital planes; *nearby OP LISL* (NOPL), which is between satellites in nearby (other than adjacent) orbital planes; and *crossing OP LISL* (COPL), which is between satellites in crossing orbital planes. The altitude of different orbital planes in a constellation is same, and satellites in these orbital planes move at the same speed. However, the direction of satellites in adjacent or nearby orbital planes is slightly different, and this leads to different relative velocities of satellites in these orbital planes. An instance of nearby OP LISLs is illustrated in Fig. 1, where satellite *x10101* has LISLs, indicated by solid yellow lines, with its nearby orbital plane neighbors *x10362* and *x12357*, and the two nearby orbital planes are marked by dashed pink lines. Similarly, Fig. 2 shows an illustration of crossing OP LISLs (with crossing OP satellites *x11232* and *x11421*) for satellite *x10101*. The dashed yellow line in these figures marks the orbital plane of satellite *x10101*.

Based on the duration of existence of LISLs between satellites, LISLs can also be classified into two types: *permanent LISLs*; and *temporary LISLs*. Intra-orbital plane LISLs are permanent, and are relatively easy to establish and maintain due to same velocities of satellites. Adjacent OP LISLs and nearby OP LISLs are usually permanent in nature but are harder to establish due to slightly different relative velocities of satellites. LISLs with certain satellites in adjacent and nearby orbital planes can also exist temporarily as we will show later. However, these temporary adjacent and nearby OP LISLs are not necessary in providing additional connectivity within the satellite network due to the existence of permanent adjacent and nearby OP LISLs.

Half of the satellites within a constellation, like Phase I of Starlink, orbit in a northeasterly direction while the other half orbit in a southeasterly direction when observed at any one region in space above the Earth [10]. Satellites in these crossing orbital planes move with high relative velocities, and crossing OP LISLs are hard to establish. These LISLs are temporary or intermittent in nature and cannot last for long durations since they are established between crossing satellites that are moving in different directions.

The satellite network that is created by establishing permanent LISLs (such as intra-OP LISLs and adjacent/nearby OP LISLs) provides a good mesh network. However, there are two separate meshes that exist within the satellite network, one between the group of satellites orbiting northeast and other between the group orbiting southeast [10]. Local connectivity between these meshes does not exist, however, it is possible to route traffic without switching between the two meshes. Creating inter-mesh links by establishing temporary crossing OP LISLs can improve routing options, and this can be critical for achieving low-latency paths within the satellite network.

## IV. Effect of Varying LISL Range on the Number of Different Types of LISLs

We study the effect of varying the range of a satellite's LISLs on the number of different types of LISLs a satellite can form with other satellites in a constellation. We define the *LISL range* of a satellite as the distance over which a satellite can establish an LISL with another satellite. The simulation is conducted using AGI's Systems Tool Kit (STK) simulator [13], and following are the main simulation parameters.

- Constellation: Starlink Phase I
- LISL range: {659; 1,319; 1,500; 1,700; 5,016} km
- Simulation period: 24 h
- Start time: 25 August 2020 16:00:00.000
- Stop time: 26 August 2020 16:00:00.000

For a specific LISL range for a satellite, we run the simulation for a period of 24 hours and observe the effect of this LISL range on the number of different types of LISLs a satellite is able to establish at this range.

We employ the satellite constellation for Phase I of Starlink for this study. This constellation is illustrated in Fig. 3. It consists of 1,584 LEO satellites and has a parameter set of {53º, 550 km, 24, 66}. This specifies an inclination of 53º, an altitude of 550 km, 24 orbital planes in the constellation, and 66 satellites in each orbital plane. The spacing between orbital planes is 15º (i.e., 360º/24), and the spacing between satellites in each orbital plane is 5.45º (i.e., 360º/66) in this figure when assuming this constellation to be uniform. The inclination of all orbital planes in the constellation is the same with reference to the Equator.

The orbital plane and satellite IDs within this constellation are generated as follows. For the 24 orbital planes, we generate the following distinct IDs: {*x101*, *x102*, *x103*, … *x124*}. For the 66 satellites within each orbital plane, we also generate distinct IDs. For example, for the satellites in the first orbital plane, we generate the following distinct IDs: {*x10101*, *x10102*, *x10103*, … *x10166*}. In this way, we generate 1,584 distinct IDs for the 1,584 satellites within the constellation.

In this study, we focus on the first satellite in the constellation, i.e., *x10101*, and study the effect of different ranges of its LISLs on the number of different types of LISLs it can establish with other satellites in the constellation. The orbital plane designation with respect to the first orbital plane carrying this satellite is as follows.

- Adjacent OPs with respect to *x101*: *x102, x124*.
- Nearby OPs with respect to *x101*: *x103, x104, ... x107, x120, x121, ... x123*.
- Crossing OPs with respect to *x101*: *x108, x109, ... x119*.

For example, when we say adjacent orbital planes of satellite *x10101*, we mean orbital planes *x102* and *x124*. Recall that half of the satellites in a constellation are orbiting in a northeasterly direction while the other half are orbiting in a southeasterly direction when observed at any one region in space above the Earth. Based on the observation of the direction of satellites in different orbital planes with respect to the first orbital plane (i.e., *x101*) in the constellation for Phase I of Starlink in Fig. 3, we come up with this designation of orbital planes with respect to *x101*.

Next, we examine the number of different types of LISLs that *x10101* is able to establish with other satellites at different LISL ranges.

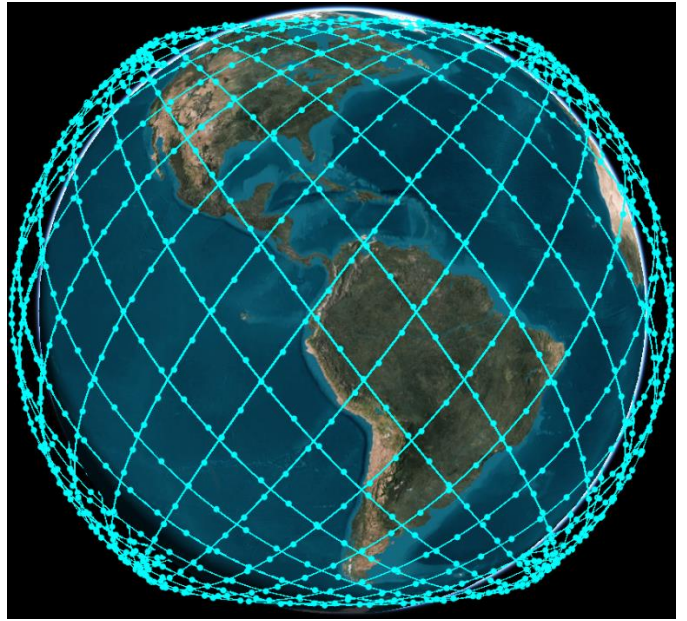

**Figure 3** *Satellite constellation for Phase I of Starlink.*

### A. Permanent LISLs

As observed from Table 1, *x10101* is able to establish only two LISLs with other satellites in the constellation at a LISL range of 659 km. These include one with its front intra-OP neighbor *x10102* and the other with its rear intra-OP neighbor *x10166*. At a LISL range of 1,500 km, *x10101* forms six LISLs as observed from Table 1, four with neighbors in the same OP and two with nearest left and right neighbors in adjacent OPs. Fig. 4 shows the scenario when *x10101* forms ten LISLs at 1,700 km range, which include four intra-OP LISLs and six adjacent OP LISLs. Satellite *x10101* and its LISLs are shown in yellow while its neighbors are shown in pink in this figure. These are instances of permanent LISLs, which means that these LISLs continue to exist during the entire 24 hour simulation period.

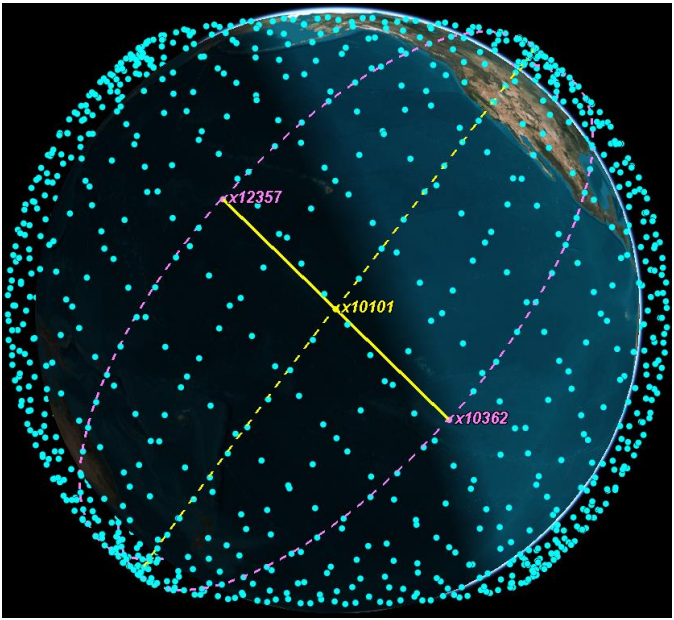

**Figure 1** *Nearby OP LISLs.*

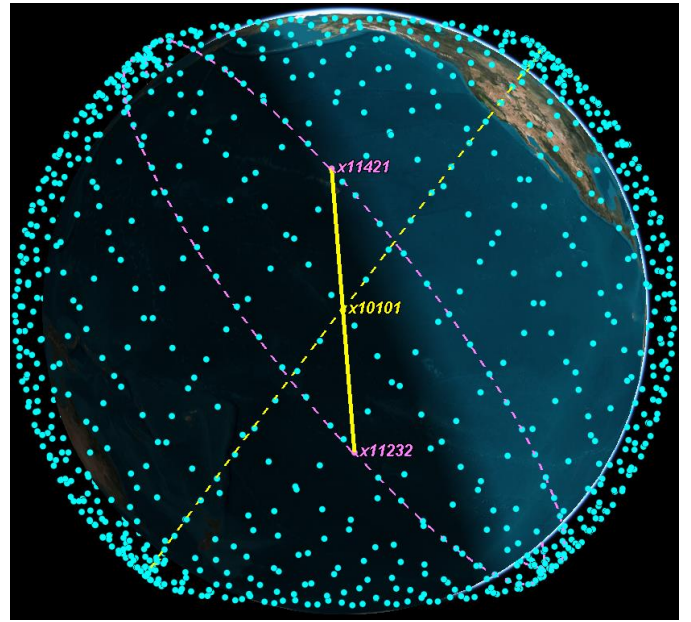

**Figure 2** *Crossing OP LISLs.*

We also consider a scenario when the range of LISLs is constrained only by visibility. For example, the maximum LISL range for an LEO satellite operating at an altitude of 550 km can be easily calculated using

$$x = \left(\sqrt{(r+h)^2 - (r+a)^2}\right) \times 2. \qquad (1)$$

In this example as shown in Fig. 5, $r$ is the radius of the Earth, $h$ is the altitude of the satellite, $a$ is the height of the atmospheric layer above the surface of the Earth, and $x$ is the maximum LISL range. Using $r = 6{,}378$ km, $h = 550$ km, and $a = 80$ km, $x$ is calculated as 5,016 km. The lowest atmospheric layer without water vapor begins at approximately 80 km above Earth's surface, which means that the minimum height of an LISL above the surface of the Earth should be 80 km. The radius of the Earth at the Equator is 6,378 km. In this figure, Earth is shown in blue, atmospheric layer in grey, and LEO satellites in Phase I of Starlink in yellow.

When there is no limit on the range of *x10101*'s LISLs other than the visibility between this satellite and other satellites in the constellation, it is interesting to note that *x10101* is able to establish 88 permanent LISLs with other satellites in its visibility (or maximum LISL) range as indicated in Table 1.

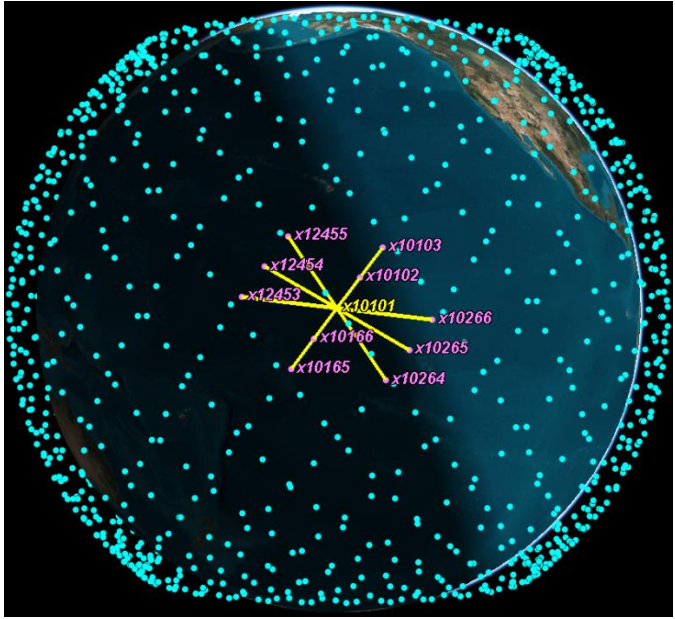

*Figure 4* Permanent LISLs at 1,700 km range.

## B. Temporary LISLs

The radius of the Earth is largest at the Equator and smallest at the Poles. Consequently, the distance between satellites in a constellation is more at the Equator and less at the Poles. For example, for a specific LISL range, *x10101* will have less satellites within its range at the Equator and more satellites within its range at the Poles. This will result in *x10101* establishing temporary LISLs with certain adjacent and nearby OP satellites near the Poles as they come within its range. Such a LISL will exist between *x10101* and its adjacent or nearby OP neighbor till that neighbor and *x10101* are within each other's LISL range near a Polar region. As *x10101* and its adjacent or nearby OP neighbor move away from the Polar region, the distance between them will increase and the LISL will cease to exist when the satellites move out of range.

The orbital period of a satellite $T$ (i.e., the time it takes a satellite to complete one circular orbit around the Earth) can be calculated using

$$T = 2\pi\sqrt{R^3/(GM_E)}, \qquad (2)$$

where $R$ is the radius of the Earth plus the altitude of the satellite from the Earth's surface, $G$ is the gravitational constant, and $M_E$ is the mass of the Earth [14]. Using $R = (6.378 \times 10^6 + 0.550 \times 10^6)$ m, $G = 6.673 \times 10^{-11}$ Nm$^2$/kg$^2$, and $M_E = 5.98 \times 10^{24}$ kg, the orbital period of a satellite at an altitude of 550 km in Starlink's Phase I constellation is calculated as 5,735.62 seconds or 1.59 hours. This means that a satellite in this constellation orbits the Earth 15 times in 24 hours.

An example of a temporary LISL with an adjacent OP neighbor (also referred to as a *temporary adjacent OP LISL* (TAOPL) in this work) is shown in Fig. 6, where *x10101* has a range of 1,700 km, and establishes an LISL with *x10263* twice near the Polar regions during its orbit around the Earth. This TAOPL is instantiated 30 times for short durations of 1,978 seconds in 24 hours as illustrated in Table 2. As seen from Table 1, there are four such TAOPLs (including the one with *x10263*) that are established for short periods between *x10101* and its four adjacent OP neighbors at 1,700 km range. Similarly, Fig. 7 shows an instance of a temporary LISL between *x10101* and its nearby OP neighbor *x10362* at 1,700 km range. We also refer to such a LISL as a *temporary nearby OP LISL* (TNOPL).

Fig. 8 shows an example of a *temporary crossing OP LISL* (TCOPL) between *x10101* and *x11232* at 1,700 km range. This temporary LISL is established between these two satellites in crossing orbital planes when they temporarily come within 1,700 km of each other while crossing each other, and this occurs 30 times for 283 second durations during a 24 hour period.

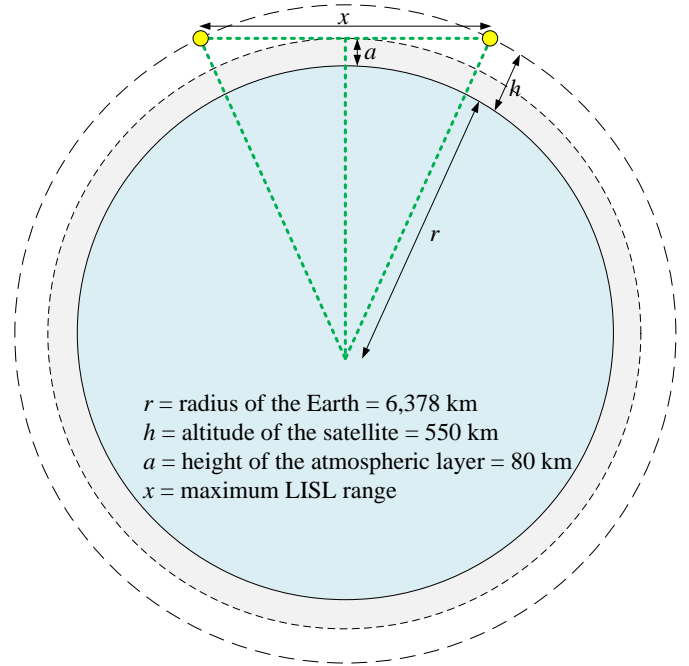

$r$ = radius of the Earth = 6,378 km
$h$ = altitude of the satellite = 550 km
$a$ = height of the atmospheric layer = 80 km
$x$ = maximum LISL range

*Figure 5* Maximum LISL range of a satellite in Phase I of Starlink.

## C. Discussion

As the LISL range of a satellite increases, it is able to establish more permanent LISLs with its intra-OP, adjacent OP, and nearby OP neighbors. At the maximum LISL range of 5,016 km, a satellite in Phase I of Starlink can establish 88 permanent LISLs with other satellites in its visibility range. Such LISLs result in creating two mesh networks within the constellation, one between the group of satellites moving northeast while the other between the group moving southeast.

A satellite is able to establish few temporary AOPLs and several temporary NOPLs at different LISL ranges. Such LISLs exist between a satellite and certain of its adjacent or nearby OP neighbors near the Poles. A satellite in Phase I of Starlink orbits the Earth 15 times in 24 hours, and a temporary AOPL or a temporary NOPL between a satellite and an adjacent OP neighbor or a nearby OP neighbor is instantiated 30 times for short durations during 24 hours. Due to permanent AOPLs and permanent NOPLs providing lasting connectivity, however, these TAOPLs or TNOPLs are not necessary in providing additional connectivity within the satellite network.

It is interesting to note that a significant number of temporary COPLs are present at all LISL ranges and this number increases with the increase in a satellite's LISL range. Such LISLs are established between two satellites in crossing orbital planes when they temporarily come within LISL range of each other while crossing each other. At the maximum LISL range, a satellite can form TCOPLs with 281 other satellites over the 24 hour period.

Local connectivity between the two mesh networks is not possible with only permanent LISLs without going halfway around the planet. Although, traffic can be routed using permanent LISLs without switching between the two meshes, the resulting routing paths may not be efficient in terms of latency. The TCOPLs create inter-mesh links and, thereby, provide an excellent opportunity to improve routing options and to achieve low-latency paths within the satellite network.

Setup times for establishing LISLs between satellites vary from a few seconds to tens of seconds [9]. These setup times and the temporary nature of crossing OP LISLs make such LISLs undesirable. Also, LISLs between satellites in crossing orbital planes suffer from challenges, like *acquisition, tracking, and pointing* (ATP) [15]. Pointing a laser beam at a moving satellite in a crossing orbital plane from another moving satellite is very difficult due to the narrow beam width of the laser beam and the different relative velocities of the satellites in crossing OPs. Therefore, a very precise ATP system is required on board a satellite platform for the laser beam emanating from one satellite to effectively connect to another satellite.

| LISL Range (km) | Number of Permanent LISLs | | | | Number of Temporary LISLs | | | |
|---|---|---|---|---|---|---|---|---|
| | Intra OP | Adjacent OPs | Nearby OPs | Total | Adjacent OPs | Nearby OPs | Crossing OPs | Total |
| 659 | 2 | 0 | 0 | 2 | 4 | 21 | 37 | 62 |
| 1,319 | 4 | 0 | 0 | 4 | 8 | 41 | 67 | 116 |
| 1,500 | 4 | 2 | 0 | 6 | 8 | 43 | 85 | 136 |
| 1,700 | 4 | 6 | 0 | 10 | 4 | 53 | 87 | 144 |
| 5,016 | 14 | 30 | 44 | 88 | 2 | 113 | 281 | 396 |

*Table 1 Effect of Different LISL Ranges on the Number of LISLs for x10101*

| Instance | Start Time | Stop Time | Duration (s) |
|---|---|---|---|
| 1 | 16:09:51.493 | 16:42:49.519 | 1978.026 |
| 2 | 16:57:39.223 | 17:30:37.247 | 1978.024 |
| 3 | 17:45:26.951 | 18:18:24.980 | 1978.029 |
| 4 | 18:33:14.681 | 19:06:12.708 | 1978.027 |
| 5 | 19:21:02.411 | 19:54:00.434 | 1978.022 |
| 6 | 20:08:50.138 | 20:41:48.163 | 1978.025 |
| 7 | 20:56:37.867 | 21:29:35.894 | 1978.027 |
| 8 | 21:44:25.598 | 22:17:23.620 | 1978.022 |
| 9 | 22:32:13.326 | 23:05:11.350 | 1978.024 |
| 10 | 23:20:01.055 | 23:52:59.080 | 1978.025 |
| 11 | 00:07:48.784 | 00:40:46.808 | 1978.023 |
| 12 | 00:55:36.512 | 01:28:34.536 | 1978.024 |
| 13 | 01:43:24.241 | 02:16:22.267 | 1978.026 |
| 14 | 02:31:11.971 | 03:04:09.993 | 1978.022 |
| 15 | 03:18:59.699 | 03:51:57.722 | 1978.023 |
| 16 | 04:06:47.428 | 04:39:45.454 | 1978.026 |
| 17 | 04:54:35.153 | 05:27:33.180 | 1978.026 |
| 18 | 05:42:22.886 | 06:15:20.910 | 1978.023 |
| 19 | 06:30:10.615 | 07:03:08.640 | 1978.026 |
| 20 | 07:17:58.340 | 07:50:56.367 | 1978.027 |
| 21 | 08:05:46.073 | 08:38:44.094 | 1978.021 |
| 22 | 08:53:33.801 | 09:26:31.827 | 1978.026 |
| 23 | 09:41:21.526 | 10:14:19.553 | 1978.027 |
| 24 | 10:29:09.258 | 11:02:07.282 | 1978.023 |
| 25 | 11:16:56.988 | 11:49:55.013 | 1978.026 |
| 26 | 12:04:44.713 | 12:37:42.740 | 1978.027 |
| 27 | 12:52:32.444 | 13:25:30.468 | 1978.024 |
| 28 | 13:40:20.174 | 14:13:18.200 | 1978.026 |
| 29 | 14:28:07.899 | 15:01:05.927 | 1978.027 |
| 30 | 15:15:55.631 | 15:48:53.655 | 1978.025 |

*Table 2 Instances of the TAOPL between x10101 and x10263*

The ATP system design in a laser terminal for LISLs involves minimizing the system's and thereby the terminal's size, weight, and power (SWaP) and cost to minimize the SWaP and cost of the satellite while providing desired accuracy. To reduce ATP system's SWaP and cost, however, its components' quality and quantity is usually reduced, which lowers its accuracy [16]. Sometimes, the complexity of the ATP system has to be reduced to reduce its SWaP and cost to meet the SWaP and cost requirements of the satellite. In other cases, a more sophisticated ATP system is required, which necessitates a higher SWaP and cost.

The ATP system in a laser terminal typically consists of a coarse pointing assembly (CPA) and a fine pointing assembly. The CPA provides coarse alignment of laser beams, and has a wide field of view and low precision while the fine pointing assembly has a narrow field of view and high precision, and aligns laser beams to achieve the desired pointing accuracy [17]. The CPA enables the ATP system to roughly point the laser beam in the desired direction without reorientation of the satellite.

Due to the low SWaP and cost requirements of small satellites like CubeSats, the ATP system design in CubeSat Lasercom Infrared CrosslinK (or CLICK) project [17] does not use a CPA and relies on satellite's attitude determination and control system for body pointing to achieve coarse pointing. Compared to CubeSats, the ATP system in SpaceX's Starlink satellites will have higher SWaP and cost. These satellites are expected to be equipped with four laser terminals to establish four simultaneous LISLs. The ATP system in each laser terminal will need a CPA for the satellite to establish LISLs with different neighbors in different directions simultaneously, and body pointing (or satellite reorientation) to reduce SWaP and cost is not feasible for Starlink satellites.

## V. Conclusions

We provided a classification of different types of LISLs that occur between satellites within a constellation to create a satellite network. Intra-OP LISLs are permanent. Adjacent and nearby OP LISLs with adjacent and nearby OP neighbors that are always within a satellite's LISL range are permanent. However, certain adjacent and nearby OP LISLs can exist temporarily near the Poles. Crossing OP LISLs are temporary in nature and cannot exist for long durations.

Using the satellite constellation for Phase I of Starlink, we studied the effect of varying a satellite's LISL range on the number of different types of LISLs. A satellite is able to establish more permanent intra-OP, adjacent OP, and nearby OP LISLs as its LISL range increases. We also observe a significant number of temporary crossing OP LISLs at all LISL ranges, and this number also increases with the increase in LISL range.

Crossing OP LISLs are currently considered undesirable due to current LISL setup times. These setup times are prohibitive and will have to be greatly reduced through technological improvement. Also, extremely efficient ATP systems having reasonable SWaP and cost will need to be developed to realize reliable temporary crossing OP LISLs in next-generation optical wireless satellite networks. Satellite reorientation for coarse pointing to reduce SWaP and cost in CubeSats is not viable for satellites in these next-generation satellite networks.

SpaceX plans to equip its Starlink satellites with four LISLs beginning in late 2020. This limits a satellite's connectivity to two satellites in the same orbital plane and to two satellites in adjacent orbital planes. Leveraging the large number of potential permanent LISLs can be beneficial in providing robust connectivity within the network. Leveraging the large number of potential temporary crossing OP LISLs can be crucial to ensure the availability of low-latency paths within the network. Consequently, satellites in next-generation optical wireless satellite networks will need to be equipped with several LISL terminals.

The impact of equipping satellites in a constellation, like Phase I of Starlink, with different number of LISL terminals on network latency is worth studying. Also, the impact of different satellite LISL ranges on network connectivity and network latency of long distance connections for data communications should be investigated. Furthermore, the effect of crossing orbital plane LISLs on network latency and the impact of different satellite altitudes in upcoming LEO/VLEO constellations on crossing orbital plane LISLs ought to be examined.

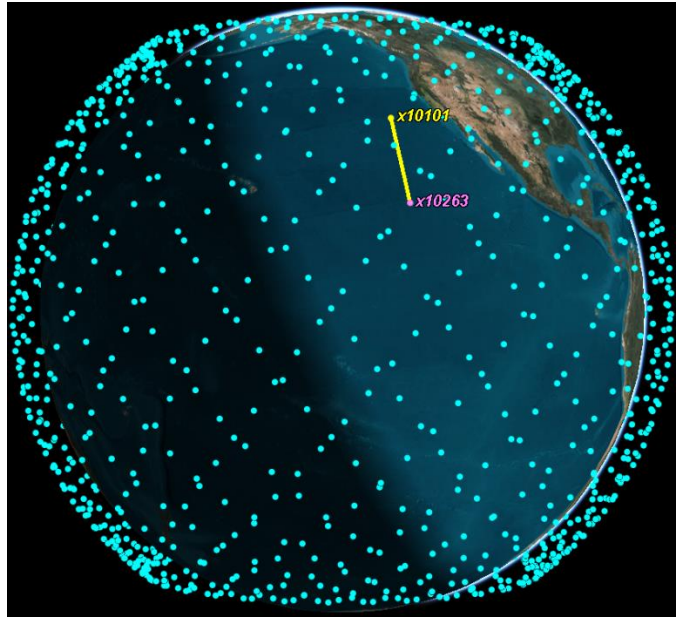

*Figure 6* An instance of an TAOPL begins as the satellites approach a Polar region.

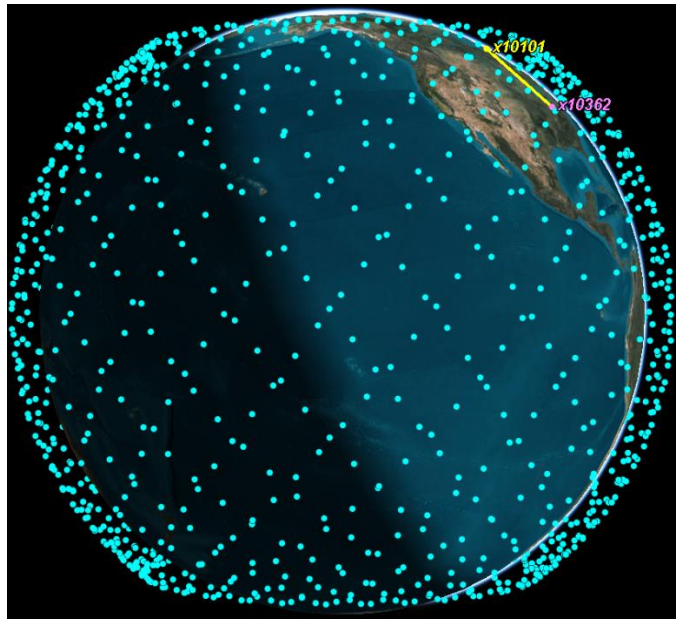

*Figure 7* An instance of an TNOPL begins near a Polar region.

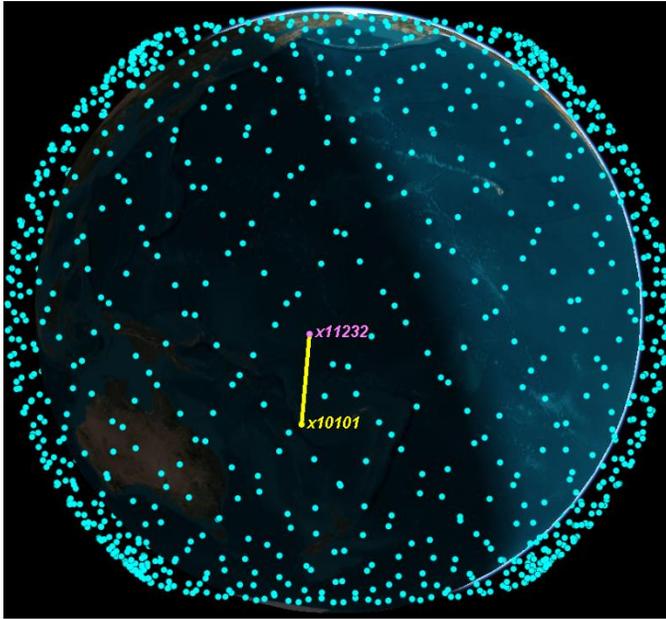

*Figure 8 An instance of an TCOPL begins as the satellites come within range of each other while crossing each other.*

## Acknowledgement

This work has been supported by the National Research Council Canada's (NRC) High Throughput Secure Networks program (CSTIP Grant #CH-HTSN-608) within the Optical Satellite Communications Canada (OSC) framework. The authors would like to thank AGI for the Systems Tool Kit (STK) platform.

## Author Information

*Aizaz U. Chaudhry (auhchaud@sce.carleton.ca)* is a Research Associate with the Department of Systems and Computer Engineering, Carleton University. His research interests include the application of machine learning and optimization in wireless networks. He is a Senior Member of the IEEE.

*Halim Yanikomeroglu (halim@sce.carleton.ca)* is a Full Professor with the Department of Systems and Computer Engineering, Carleton University. His research interests cover many aspects of wireless technologies with special emphasis on wireless networks. He is a Fellow of the IEEE.